\documentclass[
 reprint,
 showpacs, preprintnumbers,
 amsmath, amssymb,
 aps,
 pra,
 floatfix,
 superscriptaddress
]{revtex4-2}

\usepackage{amsmath}
\usepackage{graphicx}
\usepackage{dcolumn}
\usepackage{bm}
\usepackage{bbm}

\usepackage[utf8]{inputenc}
\usepackage[T1]{fontenc}
\usepackage{textgreek}
\usepackage{upgreek}
\usepackage{float}
\usepackage{hyperref}

\usepackage{newtxtext}
\usepackage{newtxmath}

\usepackage{color}
\usepackage[normalem]{ulem}
\usepackage{xcolor}
\definecolor{darkgreen}{RGB}{0 100 0}

\usepackage{natbib}

\newcommand{\beginsupplement}{%
        \setcounter{table}{0}
        \renewcommand{\thetable}{S\arabic{table}}%
        \setcounter{figure}{0}
        \renewcommand{\thefigure}{S\arabic{figure}}%
				\renewcommand{\theequation}{S.\arabic{equation}}
     }

\begin{document}

	\preprint{}

\title[phi0 vs SDE]{Link between supercurrent diode and anomalous Josephson effect revealed by gate-controlled interferometry}

\author{S.~Reinhardt}
\author{T.~Ascherl}

\affiliation{Institut f\"ur Experimentelle und Angewandte Physik, University of Regensburg, 93040 Regensburg, Germany}
\author{A.~Costa}
\affiliation{Institut f\"ur Theoretische Physik, University of Regensburg, 93040 Regensburg, Germany}

\author{J.~Berger}

\affiliation{Institut f\"ur Experimentelle und Angewandte Physik, University of Regensburg, 93040 Regensburg, Germany}
\author{S.~Gronin}
\author{G.~C.~Gardner}
\affiliation{Birck Nanotechnology Center, Purdue University, West Lafayette, Indiana 47907 USA}
\author{T.~Lindemann}
\affiliation{Birck Nanotechnology Center, Purdue University, West Lafayette, Indiana 47907 USA}
\affiliation{Department of Physics and Astronomy, Purdue University, West Lafayette, Indiana 47907 USA}

\author{M.~J.~Manfra}
\affiliation{Birck Nanotechnology Center, Purdue University, West Lafayette, Indiana 47907 USA}
\affiliation{Department of Physics and Astronomy, Purdue University, West Lafayette, Indiana 47907 USA}
\affiliation{School of Materials Engineering, Purdue University, West Lafayette, Indiana 47907 USA}
\affiliation{Elmore Family School of Electrical and Computer Engineering, Purdue University, West Lafayette, Indiana 47907 USA}


\author{J.~Fabian}
\affiliation{Institut f\"ur Theoretische Physik, University of Regensburg, 93040 Regensburg, Germany}

\author{D.~Kochan}
\affiliation{Institute of Physics, Slovak Academy of Sciences, 84511 Bratislava, Slovakia}
\affiliation{Institut f\"ur Theoretische Physik, University of Regensburg, 93040 Regensburg, Germany}

\author{C.~Strunk}
\author{N.~Paradiso}\email{nicola.paradiso@physik.uni-regensburg.de}
\affiliation{Institut f\"ur Experimentelle und Angewandte Physik, University of Regensburg, 93040 Regensburg, Germany}
%


\begin{abstract}	
In Josephson diodes the asymmetry between positive and negative current branch of the current-phase relation leads to a polarity-dependent critical current and Josephson inductance. The supercurrent nonreciprocity can be described as a consequence of the anomalous Josephson effect ---a $\varphi_0$-shift of the current-phase relation--- in multichannel ballistic junctions with strong spin-orbit interaction. In this work, we simultaneously investigate $\varphi_0$-shift and supercurrent diode efficiency on the same Josephson junction by means of a superconducting quantum interferometer. By electrostatic gating, we reveal a direct link between $\varphi_0$-shift and diode effect. 
Our findings show that the supercurrent diode effect mainly results from magnetochiral anisotropy induced by spin-orbit interaction in combination with a Zeeman field.


\end{abstract}

\maketitle

\section{Introduction}

In solids, spin-orbit interaction (SOI) makes it possible to control orbital degrees of freedom by acting on the electron spin, and vice-versa~\cite{Winkler2003spin,Manchon2015}. In superconductors~\cite{Smidman2017}, the impact of SOI can be particularly spectacular, since it enables phenomena which go beyond the realm of conventional $s$-wave superconductors, as e.g.~topological phases~\cite{Sato2017}, finite-momentum superconductivity~\cite{YuanFuFM,Hart2017,ChenNatComm2018}, Lifshitz invariant~\cite{Edelstein1996,FuchsPRX,Kochan2023diode}, Ising superconductivity~\cite{LuIsing2015}, anomalous Josephson effect~\cite{Buzdin2008,Reynoso2008,Yokoyama2014,Szombati2016,Strambini2020} and intrinsic supercurrent diode effect~\cite{Ando2020,Baumgartner2022,Wu2022,DiezMerida2023,Lin2022,Turini2022,Pal2022,Jeon2022,Lotfizadeh2023,Mazur2022,Costa2023sign,Bauriedl2022,YunPRR2023,Banerjee2023phase,Sundaresh2023}. In what follows, we shall focus on the last two effects and on their relation in Josephson junctions.

The anomalous Josephson effect manifests itself in a  phase offset $\varphi_0$ at zero current, $I(\varphi_0)=0$, in the current-phase relation (CPR)~\cite{Szombati2016,Assouline2019,Mayer2020b,Dartiailh2021,Strambini2020,FrolovQW2022,Haxell2023}. This also implies a
finite supercurrent at zero phase difference $I(\varphi=0)\neq 0$.  The effect requires the simultaneous breaking of both inversion and time-reversal symmetry~\cite{Krive2004}, which can be provided by SOI and Zeeman field, respectively.

The same symmetries need to be broken in order to observe the supercurrent diode effect (SDE), namely, the dependence of the critical current on the bias polarity. This effect can be trivially obtained, e.g., in asymmetric superconducting quantum interference devices 
(SQUIDs)~\cite{barone} or, more generally, when in a film an inhomogeneous supercurrent distribution is coupled to a flux. Recently, it was shown~\cite{Ando2020,Baumgartner2022,Wu2022,Turini2022,Pal2022,Jeon2022,Lotfizadeh2023,Mazur2022,Costa2023sign,Banerjee2023phase,Sundaresh2023} that supercurrent rectification can as well emerge as an intrinsic feature of \textit{homogeneous} quasi-2D systems subjected to a Zeeman field. Such nontrivial SDE is a new precious probe of the condensate physics (in films)~\cite{Kochan2023diode} and of Andreev bound states (ABSs) in Josephson junctions~\cite{Costa2023sign,Banerjee2023phase}, including possible topological properties~\cite{Legg2023,Pientka2017,Scharf2019}.

Several mechanisms have been proposed to explain such intrinsic SDE in films~\cite{daido2021prl,yuan2021pnas,he2021njp,Ilic2022,Kochan2023diode,HuPRL2023}
and Josephson junctions~\cite{Grein2009,Zhang2022,scammell2022theory,Davydova2022,Fu2022TopoDiode,BoLu2022,costa2023microscopic}. 
In superconducting-normal-superconducting (SNS) junctions the supercurrent can be computed in terms of the ABSs in the N weak link. In experiments, ABSs can be directly probed by tunnel spectroscopy~\cite{Fornieri2019,Banerjee2022,Haxell2023}, while their effect on the CPR can be deduced from IV-characteristics, inductance versus current measurements~\cite{Baumgartner2022,BaumgartnerSI2022,Costa2023sign} and SQUID experiments~\cite{NichelePRL2020}.

Early observations on Josephson diodes were interpreted in terms of  $\varphi_0$-shift in ballistic systems with skewed CPR~\cite{Baumgartner2022,BaumgartnerSI2022,Jeon2022,Pal2022,Turini2022}. Within this picture, the SDE ultimately originates (as it does the $\varphi_0$-shift) from SOI.
An alternative model, proposed by Banerjee \textit{et al.}~\cite{Banerjee2023phase} based on the theory of Ref.~\cite{Davydova2022}, explains the same effect in terms of a purely orbital mechanism. To date, it is not clear yet to which extent the two mechanisms (namely, the SOI-based and the purely orbital mechanism) contribute to the supercurrent rectification observed in experiments.

\begin{figure*}[tb]
\centering
\includegraphics[width=\textwidth]{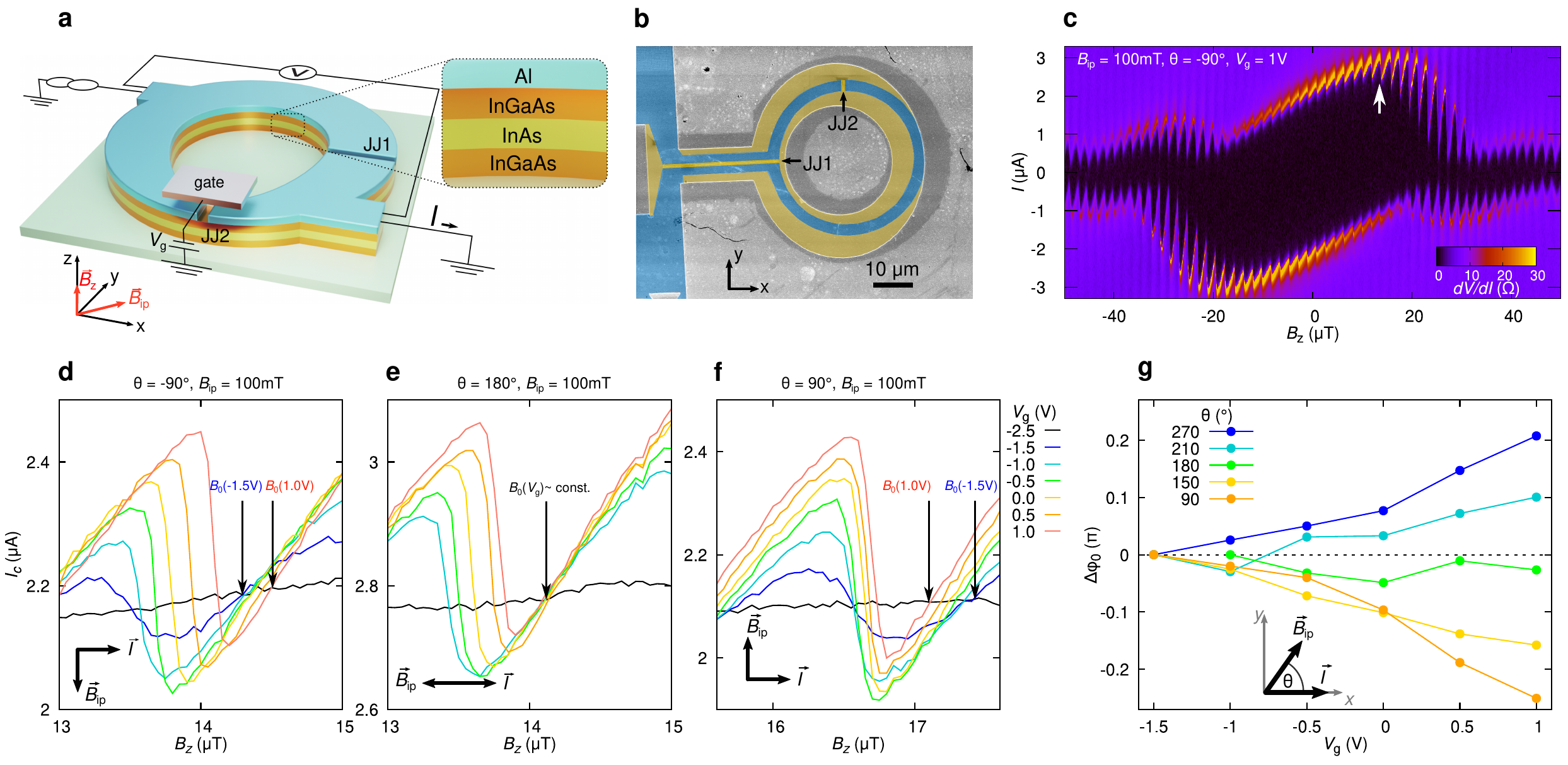}
\caption{\textbf{Asymmetric SQUID device with reference junction and gate-controllable $\varphi_0$ junction.} \textbf{a}, (Left)~Schematic illustration of the device. The SQUID consists of a large reference junction (JJ1) and of a small junction (JJ2) which is coupled to a gate. By applying a magnetic field along $\hat{y}$, a $\varphi_0$-shift is induced in the CPR of the latter junction, which can be controlled by electrostatic gating. (Right)~Scheme of the topmost layers of the heterostructure. The black arrow indicates the direction of the positive current bias $I$.
\textbf{b}, False-color scanning electron microscopy image of the device taken before gate patterning. The pristine superconducting Al/InAs leads are highlighted in turquoise, the areas where Al is selectively etched in yellow (including the weak links, highlighted by the black arrows). The remaining parts in gray correspond to deeply etched regions, where both the Al film and the topmost semiconducting layers are etched. 
\textbf{c}, The color plot shows the SQUID differential resistance versus out-of-plane field $B_z$ and current $I$,  for $\theta=-90^{\circ}$ and $B_{\text{ip}}=100$~mT (i.e., $B_x=0$, $B_y=-100$~mT) at $T=40$~mK. The white arrow indicates where anomalous phase shifts were measured (see panels \textbf{d-f}). 
\textbf{d}, SQUID critical current $I_c$ as a function of $B_z$, for $\theta$ and $B_{\text{ip}}$ as in \textbf{c}. The different curves refer to different gate voltages $V_g$. We define $B_0(V_g)$ as the crossing of each curve with the $V_g=-2.5$~V reference curve (black).
\textbf{e} and \textbf{f}, Corresponding measurements for the same $B_{\text{ip}}=100$~mT but, respectively, $\theta=180^{\circ}$ and $\theta=90^{\circ}$.
\textbf{g}, Plot of $\Delta \varphi_0 (V_g) \equiv 2\pi A[B_0(V_g)-B_0(V_g=-1.5\text{~V})]/ \Phi_0$, for different $\vec{B}_{\text{ip}}$ orientations, i.e., for different $\theta$. Here, $A$ is the loop area and $\Phi_0$ the flux quantum.}
\label{fig:Fig1SR}
\end{figure*}

In this work, we make use of an asymmetric SQUID with mutually orthogonal junctions to directly measure both the anomalous $\varphi_0$-shift and the SDE on the same junction. By gating, we can electrostatically control both effects and highlight their relation. Finally, by measuring the temperature dependence of the $\varphi_0$-shift and of the diode efficiency we demonstrate that the former is a necessary but not sufficient condition for the SDE.  This latter requires in fact the presence of higher harmonics in the CPR, which are quickly suppressed by increasing the temperature. We comment on our results in light of alternative models proposed in the literature and compare the temperature dependence of $ \varphi_0 $ and the SDE to the predictions of a minimal theoretical model.

\section{Device description and experimental results}

Figure~\ref{fig:Fig1SR}\textbf{a} shows a scheme of our SQUID. 
The device is fabricated starting from a molecular beam epitaxy-grown heterostructure featuring an InGaAs/InAs/InGaAs quantum well, capped by a 5~nm-thick epitaxial Al film~\cite{Shabani2016,Wickramasinghe2018, baumgartner2020,Baumgartner2022}. The quantum well 
hosts a 2D electron gas (2DEG) with a proximity-induced superconducting gap inherited from the Al film. By deep wet etching, we define an asymmetric SQUID loop. The actual geometry is shown in the false-color scanning electron microscopy image in Fig.~\ref{fig:Fig1SR}\textbf{b}, where the turquoise areas indicate the pristine Al/InGaAs/InAs/InGaAs regions, while the gray areas refer to deeply etched (insulating) regions. To obtain the two normal (N) weak links, we selectively etch the Al film (yellow areas in Fig.~\ref{fig:Fig1SR}\textbf{b}). The reference Josephson junction 1 (JJ1) is 28~\textmu m-wide and 120~nm-long, whereas the Josephson junction 2 (JJ2) is 2.7~\textmu m-wide and 100~nm-long. Finally, a gate is fabricated on top of JJ2, which allows us to control the electron density in the N-link and thus the critical current $I_{c,2}$ of this junction.
The two junctions are mutually perpendicular, so that an in-plane magnetic field $\vec{B}_{\text{ip}}$ parallel to the short junction, i.e., along $\hat{y}$, (see Fig.~\ref{fig:Fig1SR}\textbf{a,b}) will induce magnetochiral effects~\cite{Baumgartner2022} in the short junction only, and not in the reference junction. 
Here, we take as positive $\hat{z}$ direction that  perpendicular to the 2DEG and directed from the substrate towards the Al (see Fig.~\ref{fig:Fig1SR}\textbf{a}), which corresponds to the direction opposite 
to the built-in electric field in the quantum well~\cite{Baumgartner2022,ZhangPRM2023} which provides 
SOI in the 2DEG~\cite{Baumgartner2022,ZhangPRM2023}.

We measure differential resistance in a 4-terminal geometry as a function of DC current.
In what follows, we indicate as $I_c$ the (measured) SQUID critical current (see Methods) and as $I_{c,i}$ ($i=1,2$) the (deduced) critical current in junction $i$. When needed, we use the superscript $+$ ($-$) to indicate positive (negative) current from source to drain (see Fig.~\ref{fig:Fig1SR}\textbf{a}), i.e., supercurrent in the positive (negative) $x$ direction in JJ2. 
For the in-plane field, we interchangeably use either Cartesian components, or 
magnitude and angle parametrization, i.e.~$\vec{B}_{\text{ip}}=B_x\hat{x}+B_y\hat{y}=B_{\text{ip}}(\cos\theta,\sin\theta)$, see cartesian axes in Fig.~\ref{fig:Fig1SR}\textbf{a}.

Figure~\ref{fig:Fig1SR}\textbf{c} shows the  color plot of the differential resistance versus out-of-plane field $B_z$ and DC current $I$, measured with an applied field $B_y=-100$~mT at $T=40$~mK. The fast oscillations have period $1.8$~\textmu T, corresponding to a flux quantum $\Phi_0=h/2e$ applied to the loop. Such oscillations are superimposed to the Fraunhofer pattern of the reference junction, whose central and first side lobes are visible. The plot displays an evident asymmetry around the $B_z=0$ and $I=0$ axes, while it is approximately point-inversion-symmetric around the origin, namely $I_c(B_z)\approx -I_c(-B_z)$. As discussed in the Supplementary Information, such asymmetry is due to screening effects, which arise when there are current-dependent corrections to the fundamental SQUID relation
\begin{equation}
    \gamma_1-\gamma_2=\frac{2\pi}{\Phi_0}\Phi,
    \label{eq:squid}
\end{equation}
where $\gamma_i$ it the gauge-invariant phase drop at the $i$-th junction, $\Phi_0$ is the flux quantum and $\Phi$ the flux through the loop. The corrections originate either from the additional flux induced by the SQUID current itself, or by the phase drop accumulated along the loop arms~\cite{barone,Tinkhambook}. The former correction is proportional to the geometric loop inductance, while the latter is proportional to the kinetic inductance of the SQUID arms~\cite{Goswami2016}.   
As discussed in the Methods, in our sample, screening is mostly dominated by the large kinetic inductance of the thin Al film,  whose sheet inductance is $L_\square \sim 30$~pH. The geometric loop inductance is comparatively small, see Supplementary Information. As shall be discussed below, screening effects hinder the measurement of both the absolute value of the anomalous shift $\varphi_0$ and that of the diode efficiency. 
Indeed, the determination of the absolute $\varphi_0$ is challenging even in SQUID devices with low screening and reference devices~\cite{Haxell2023,Assouline2019}. One of the difficulties is the fact that an accurate determination of $\varphi_0$ requires a reproducible $B_z$ control on the microtesla scale, while an in-plane field of the order of tens or hundreds of millitesla is swept. 

\begin{figure*}[htb]
\centering
\includegraphics[width=\textwidth]{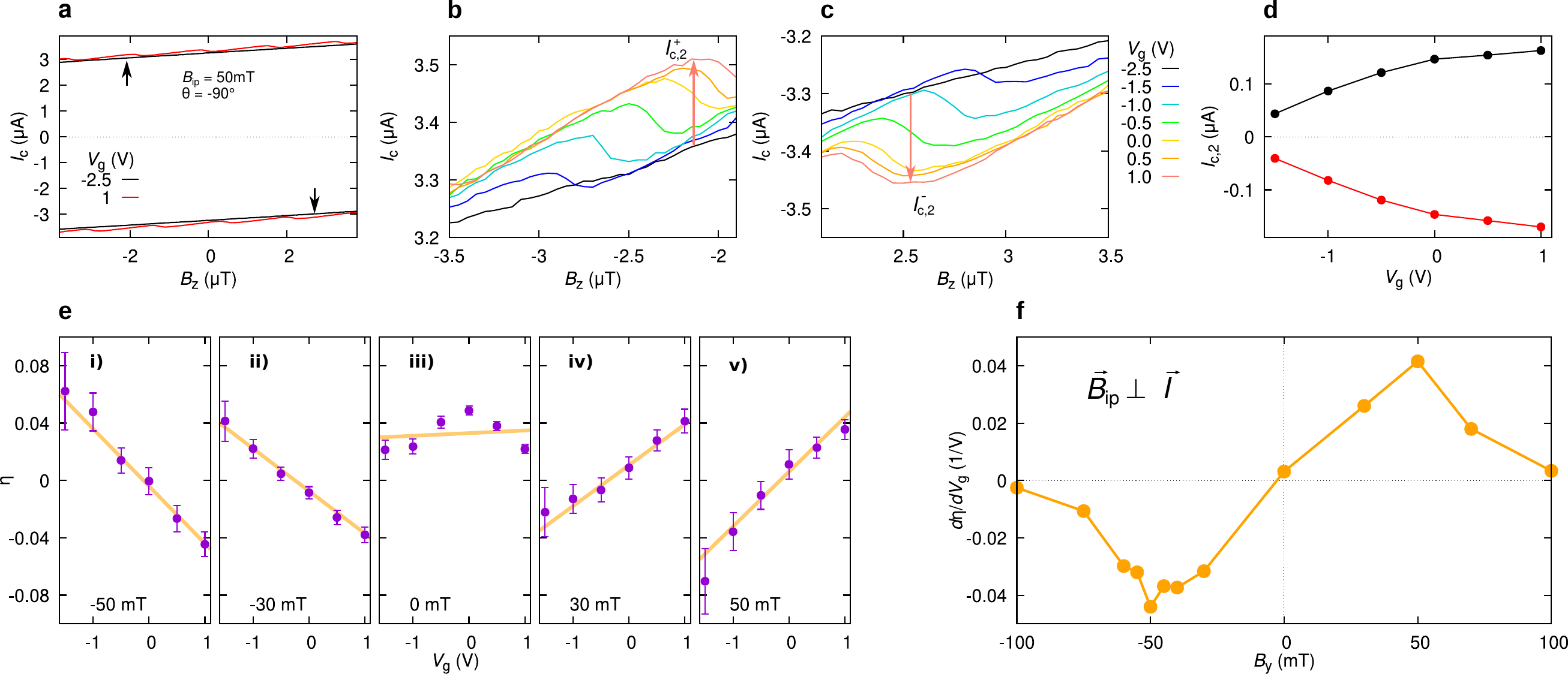}
\caption{
\textbf{Gate control of the supercurrent diode effect.} 
\textbf{a}, Positive and negative SQUID critical current for an out-of-plane field $B_z$ close to zero, $T=40$~mK, and $B_y=-50$~mT. The red (black) curve refers to a gate voltage $V_g=1.0$~V ($V_g=-2.5$~V). The upper and lower arrow indicates the oscillation highlighted in panel \textbf{b} and \textbf{c}, respectively. They are located asymmetrically in $B_z$, eliminating the trivial diode effect of the background.
\textbf{b}, Positive critical current $I_c^{+}$ versus out-of-plane field $B_z$. The red arrow indicates $I_{c,2}^+$ for the $V_g=1$~V-curve, see text. 
\textbf{c}, Negative critical current $I_c^{-}$ versus $B_z$. The red arrow indicates $I_{c,2}^+$ for the $V_g=1$~V-curve.
The different curves in \textbf{b} and \textbf{c} correspond to different values of the gate voltage $V_g$. Measurements are performed at $T=40$~mK and $B_y=-50$~mT.
\textbf{d}, $I_{c,2}^{+}$ and $I_{c,2}^{-}$ versus $V_g$ for $B_y=-50$~mT. 
\textbf{e}, Diode efficiency $\eta \equiv 2 (I_c^+-|I_c^-|)/(I_c^++|I_c^-|)$ versus $V_g$, for $B_y=-50$~mT (i), -30~mT (ii), 0~mT (iii), 30~mT (iv), 50~mT (v). 
\textbf{f}, Slope of the diode efficiency $d\eta/dV_g$, plotted versus $B_y$. 
}  
\label{fig:Fig2SR}
\end{figure*}

The determination of the \textit{absolute} $\varphi_0$ is usually difficult in systems with large kinetic inductance. Similar to Refs.~\cite{Mayer2020b,Dartiailh2021,Haxell2023}, we shall measure the \textit{relative} shift of the CPR  with respect to that for large negative gate voltage.
Figure~\ref{fig:Fig1SR}\textbf{d} shows SQUID  oscillations measured for different gate voltages $V_g$. The measurement is performed with an in-plane field $B_{\text{ip}}=100$~mT applied perpendicular to JJ2 ($\theta=-90^{\circ}$) at $T=40$~mK. The oscillations are taken near the maximum of the tilted Fraunhofer pattern, indicated by the upper arrow in Fig.~\ref{fig:Fig1SR}\textbf{c}. The black curve ($V_g=-2.5$~V) refers to a completely pinched-off JJ2 ($I_{c,2}=0$), and serves as zero-current baseline. The first curve for which oscillations are clearly visible is that for $V_g=-1.5$~V. We shall label as $B_0(V_g)$ the crossing with positive slope of each $I_c(V_g)$ curve with the baseline $I_c(V_g=-2.5\text{~V})$, see Fig.~\ref{fig:Fig1SR}\textbf{d}. As discussed in the Supplementary Information, at these crossings the supercurrent $I_2$ in JJ2 vanishes, therefore its gauge invariant phase difference is, by definition, the anomalous shift $\varphi_0$. 
We consider variations of $\varphi_0$ with respect to the reference voltage $V_g=-1.5$~V~\cite{Haxell2023}, namely,   $\Delta \varphi_0  \equiv 2\pi A_\text{loop} [B_0(V_g)-B_0(V_g=\text{-1.5~V})]/\Phi_0$, where $A_\text{loop} = 1150$~\textmu m$^2$ is the loop area.


Figures~\ref{fig:Fig1SR}\textbf{e} and~\ref{fig:Fig1SR}\textbf{f} show the results of the same measurements after two subsequent -90$^{\circ}$ sample rotation, namely, for $\theta=180^{\circ}$ and  $\theta=90^{\circ}$, respectively. From Fig.~\ref{fig:Fig1SR}\textbf{e} we deduce that for $\vec{B}_{\text{ip}}\parallel \vec{I}$, the curves crosses the baseline with positive slope nearly at the same $B_z$ (i.e., $B_0(V_g)$ is constant). Instead, for $\vec{B}_{\text{ip}}\perp \vec{I}$, $B_0(V_g)$ monotonically increases (decreases) with $V_g$ for negative (positive) sign of $\hat{e}_z\cdot (\vec{B}_{\text{ip}}\times \vec{I})$. This is a clear signature of the magnetochiral nature of the anomalous Josephson effect~\cite{Baumgartner2022}.

The variation of $\Delta \varphi_0 (V_g)$  for different $\vec{B}_{\text{ip}}$ orientations (i.e., for different $\theta$ with $|\vec{B}_{\text{ip}}|=100$~mT) is plotted in Fig.~\ref{fig:Fig1SR}\textbf{g}.
We stress that, since we subtract $\varphi_0(V_g=\text{-1.5V})$ (as in the definition of $\Delta \varphi_0$), what is important in Fig.~\ref{fig:Fig1SR}\textbf{g} is the monotonic increase or decrease of $\Delta \varphi_0$ with the gate voltage. The graph clearly shows the proportionality of $\Delta \varphi_0$ to~\cite{Baumgartner2022} $-\hat{e}_z\cdot (\vec{B}_{\text{ip}}\times \vec{I})$, as expected by SOI-based models for the anomalous Josephson effect~\cite{Buzdin2008}.
For all curves, the magnitude of $|\Delta \varphi_0|$ increases with $V_g$. The monotonic increase is expected, since in this type of InGaAs/InAs/InGaAs quantum wells, a positive gate voltage increases the built-in electric field~\cite{ZhangPRM2023} responsible for the Rashba SOI~\cite{Baumgartner2022,ZhangPRM2023}.

The main goal of our experiments is to establish a relation between the anomalous $\varphi_0$-shift and intrinsic SDE by measuring both phenomena \textit{on the same junction}. For this purpose, we investigate SQUID oscillations for both current bias polarities in order to deduce both the positive ($I_{c,2}^+$) and the negative ($I_{c,2}^-$) critical current of JJ2. 
Figure~\ref{fig:Fig2SR}\textbf{a} shows the SQUID interference pattern measured in the vicinity of $B_z=0$ for $B_y=-50$~mT at $V_g=1.0$~V (red), together with the reference curve at $V_g=-2.5$~V (black, where $I_{c,2}=0$ and SQUID oscillations vanish).
Figure~\ref{fig:Fig2SR}\textbf{b} shows several $I_c^+(B_z)$ curves for different gate voltages $V_g$, measured at $B_y=-50$~mT and $T=40$~mK, as in Fig.~\ref{fig:Fig2SR}\textbf{a}. We focus on one particular oscillation highlighted by the upper arrow in Fig.~\ref{fig:Fig2SR}\textbf{a}.
The corresponding $I_c^-$ curves for the opposite $B_z$ range (lower arrow in Fig.~\ref{fig:Fig2SR}\textbf{a}) are shown in panel \textbf{c}. We use the baseline curve in Fig.~\ref{fig:Fig2SR}\textbf{a} (black, $V_g=-2.5$~V)  as a reference to extract $I_{c,2}^+$ from data in Fig.~\ref{fig:Fig2SR}\textbf{b} and  $I_{c,2}^-$ from Fig.~\ref{fig:Fig2SR}\textbf{c}. For each $V_g$, $I_{c,2}^+$ corresponds to the maximum in $B_z$ (see red arrow in Fig.~\ref{fig:Fig2SR}\textbf{b}) of the difference $I^+_c(B_z,V_g)-I^+_c(B_z,V_g=\text{-2.5~V})$. $I_{c,2}^-$ is deduced in a similar way from $I^-_c(B_z,V_g)-I^-_c(B_z,V_g=\text{-2.5~V})$. 


The resulting $I_{c,2}^+(V_g)$ and $I_{c,2}^-(V_g)$ are plotted in Fig.~\ref{fig:Fig2SR}\textbf{d}. As a figure of merit for the supercurrent rectification, we use the supercurrent diode efficiency $\eta \equiv (I_{c,2}^+-|I_{c,2}^-|)/\langle I_{c,2} \rangle$, with $\langle I_{c,2} \rangle\equiv (I_{c,2}^++|I_{c,2}^-|)/2 $. The  efficiency $\eta(V_g)$ is plotted in Fig.~\ref{fig:Fig2SR}\textbf{e}  for different values of $B_y$ in the different subpanels (i-v).  As explained in the Supplementary Information, owing to SQUID screening effects  $\eta$ is affected by trivial offset, which means that data in Figs.~\ref{fig:Fig2SR}\textbf{e} might be subjected to a spurious vertical shift. 
Nevertheless, the \textit{variation} of $\eta$ when $V_g$ is swept remains unaffected by screening artifacts. Since the modulation of $\eta$ with $V_g$ is roughly proportional to $\eta$ itself, we  use $\partial \eta /\partial V_g$ as a measure of $\eta$. In particular, we expect that the slope $\partial \eta /\partial V_g$ must display the same magnetochiral behavior as $\eta$ itself. 
Data in Fig.~\ref{fig:Fig2SR}\textbf{e}(i-v) show that $\partial \eta /\partial V_g$ is linear in $B_y$, similar to the rectification efficiency $\eta$ reported in single Josephson junctions from similar materials~\cite{Baumgartner2022,Jeon2022,Lotfizadeh2023,Turini2022}.
The complete overview of the $B_y$-dependence of $\partial \eta /\partial V_g$ is shown in Fig.~\ref{fig:Fig2SR}\textbf{f}. As in previous reports on Josephson diodes, the rectification efficiency is linear in $B_y$ only up to a certain threshold (here at $|B_y|=50$~mT): after that, a clear suppression is observed~\cite{Costa2023sign}.

\begin{figure*}[htb]
\centering
\includegraphics[width=\textwidth]{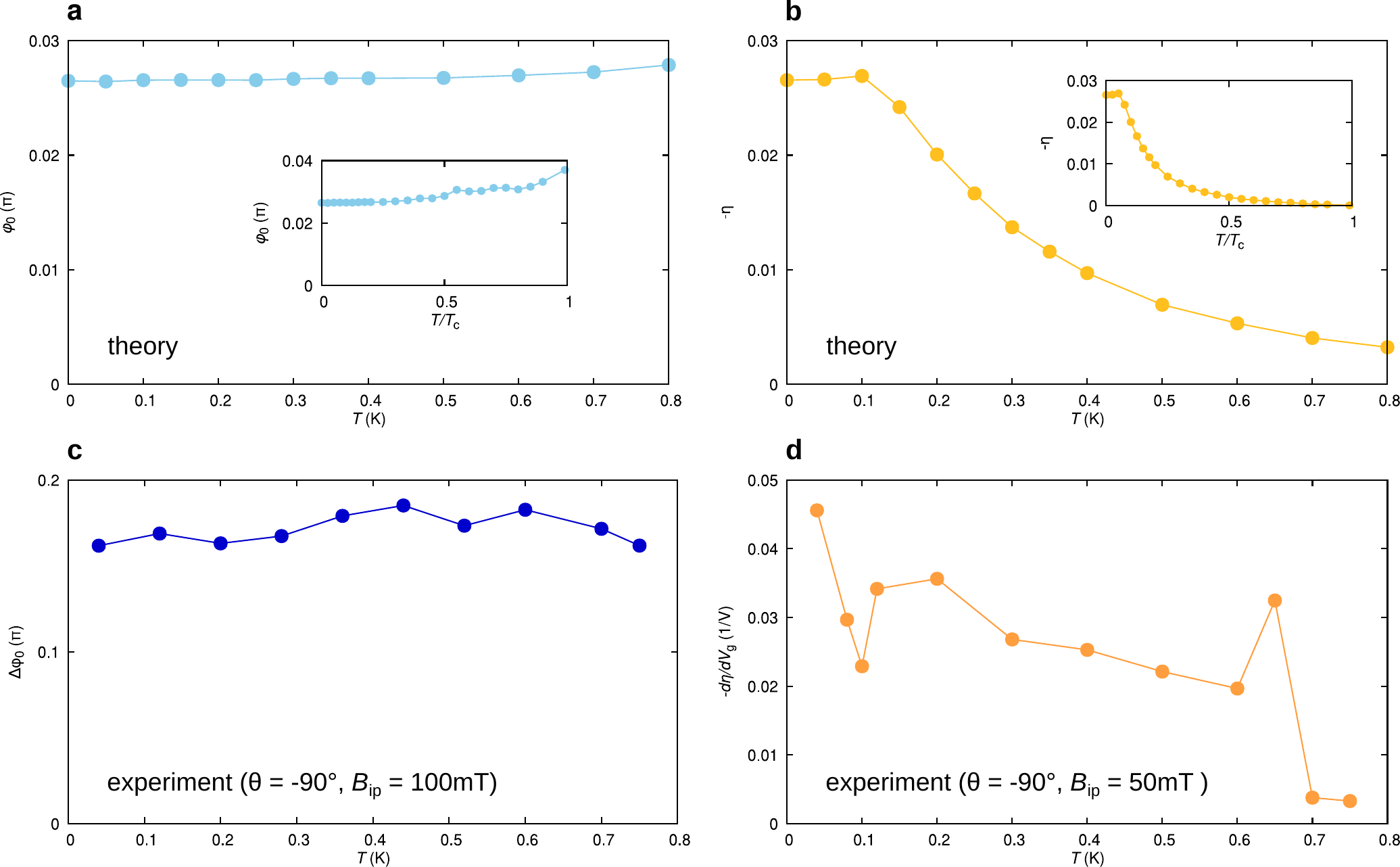}
\caption{
\textbf{Temperature dependence of  $\varphi_0$ and $\eta$:  theory and experiment.} 
\textbf{a}, Computed temperature dependence of the (sign changed) anomalous phase shift $\varphi_0$ for a Zeeman parameter of~$ \lambda_\mathrm{Z} = -0.40 $~(which would correspond to $B_y = -100$~mT for $g $-factor $|g|=12$). We assume $T_c=2.0$~K as in the experiments. Inset: full-range $\varphi_0(T)$ graph plotted for $T$ up to $T=T_c$.  
\textbf{b}, Computed temperature dependence of the supercurrent diode rectification efficiency $-\eta$ for the same parameters as in \textbf{a}. Inset: full-range $-\eta(T)$ graph up to $T=T_c$. 
\textbf{c}, Measured $\Delta \varphi_0$ (as defined in the text) at $V_g=1.0$~V and $B_y=-100$~mT, plotted as a function of temperature.
\textbf{d}, Temperature dependence  of $d\varphi_0/dV_g$ (as defined in the text) at $B_y=50$~mT. For ease of comparison, all main graphs are plotted up to $T=0.8$~K, the highest temperature for which $\Delta \varphi_0$ and $d\varphi_0/dV_g$ was measurable.
}
\label{fig:Fig3SR}
\end{figure*}

The results shown so far provide strong evidence of the link between anomalous phase shift and SDE. Both effects are linear in $B_y$. Both $\Delta \varphi_0$ and $\eta$ can be modulated by a gate voltage in the accessible range~$V_g \in $[-1.5~V, 1.0~V].  This is consistent with the SOI-based model~\cite{Baumgartner2022,Jeon2022,BaumgartnerSI2022,Kochan2023diode,costa2023microscopic} of the SDE, where the supercurrent rectification is accompanied by the anomalous Josephson shift $\varphi_0$. The $\varphi_0$-shift is a necessary but not sufficient condition for the supercurrent nonreciprocity. To break the symmetry between positive and negative part of the CPR, multiple ballistic channels are needed~\cite{Yokoyama2014,Costa2023sign}. In finite-width junctions with parabolic dispersion, each $i$-channel has a different $\varphi_{0,i}$-shift, owing to the different Fermi velocity $v_{{F,i}}$: in fact, for a single ballistic channel, $\varphi_{0,i} \propto v_{F,i}^{-2}$~\cite{Buzdin2008}.
The total CPR is the sum of all single-channel CPR contributions $I_i(\varphi + \varphi_{0,i})$. The sum of skewed CPRs with different $\varphi_{0,i}$-shift leads to an asymmetric total CPR, whose positive and negative branch are different. The skewness (i.e., the content of higher harmonics) of the individual CPRs is crucial, since otherwise the sum of sinusoidal CPRs would  always lead to a sinusoidal --i.e., reciprocal-- CPR. 
The relation between $\eta$ and $\varphi_0$ is evident from the comparison between Fig.~\ref{fig:Fig1SR}\textbf{g} and Fig.~\ref{fig:Fig2SR}\textbf{e}.  
On the other hand, since the SDE also relies on the presence of higher harmonics in the CPR~\cite{Baumgartner2022}, we expect that $\eta$ (and thus $\partial \eta /\partial V_g$) will be highly sensitive to the junction transparency and temperature, as opposed to the anomalous shift $\varphi_0$.

Figures~\ref{fig:Fig3SR}\textbf{a} and~\ref{fig:Fig3SR}\textbf{b} respectively show the temperature dependence of $\varphi_0$ and $\eta$ as obtained from a simple theoretical model~(described in Methods and in Refs.~\cite{Costa2023sign,costa2023microscopic}) for a Josephson junction in the short-ballistic limit with Zeeman interaction. The junction separates two semi-infinite superconducting 2DEGs with zero-temperature superconducting gap $\Delta^*(0) \approx 130 \, \upmu \mathrm{eV} $~\cite{baumgartner2020}, and critical temperature~$ T_\mathrm{c} \approx 2 \, \mathrm{K} $. We note that the anomalous shift $\varphi_0$ is nearly $T$-independent, whereas the supercurrent rectification is strongly suppressed already for $T>100$~mK.

In Fig.~\ref{fig:Fig3SR}\textbf{c} and \ref{fig:Fig3SR}\textbf{d} we show the measured temperature dependence of $\Delta \varphi_0$ and that of $\partial \eta /\partial V_g$, which we take as measures of $\varphi_0 (T)$ and $\eta(T)$, respectively. $ \Delta \varphi_0(T)$ is measured at $B_y=-100$~mT, ($B_{\text{ip}}=100$~mT, $\theta=-90^{\circ}$), where the field magnitude is set large enough to limit screening effects to an acceptable level. Instead, $\partial \eta /\partial V_g (T)$  is measured at $B_y=50$~mT, where the SDE is maximal, see Fig.~\ref{fig:Fig2SR}\textbf{f}. Figure~\ref{fig:Fig3SR}\textbf{c} shows that $\Delta \varphi_0$ is temperature-independent within the experimental accuracy. In contrast, the (gate modulation of the) supercurrent rectification is clearly suppressed already at temperatures well below $T_c$, as shown in Fig.~\ref{fig:Fig3SR}\textbf{d}. Both observations match the corresponding theory predictions. 

A comment is in order about the sign and magnitude of the effects. Both our experimental data and analytical model show that if the Rashba SOI-inducing electric field is directed along $-\hat{z}$ (as in Ref.~\cite{ZhangPRM2023} and in the Supplementary Information of Ref.~\cite{Baumgartner2022}), $\vec{B}_{\text{ip}}$ along $+\hat{y}$, and the positive current bias along $+\hat{x}$, then $\varphi_0<0$ [where $I(\varphi_0)=0$, $\partial_{\varphi}I(\varphi_0)>0$] and $\eta>0$. Instead, the magnitude of $\varphi_0$ and $\eta$ predicted by ballistic theory~\cite{Buzdin2008} is smaller than the one measured in our and in other experiments in the literature~\cite{Mayer2020b,Assouline2019}. A possible explanation for this discrepancy could be disorder in and near the  junction~\cite{Assouline2019}, since diffusive models predict a much larger $\varphi_0$. Another possibility is 
the enhancement of SOI due to interaction of quantum well electrons with the image charges that are formed in a nearby Al gate. The nontrivial property of the induced image-potential
that depends on the electron density of 2DEG gives a feedback on SOI that superimposes with the 
innate Rashba SOI of the quantum well without metallic gate as demonstrated in Refs.~\cite{McLaughlan_2004,PhysRevB.95.045138,PhysRevB.94.115412}.

\section{Discussion}
The main goal of our study is to elucidate the physical mechanism behind the intrinsic SDE in single, homogeneous Josephson junctions. The effect has been so far explained by two different models: one~\cite{Yokoyama2014,Baumgartner2022,BaumgartnerSI2022} is based on the combination of Rashba SOI plus Zeeman interaction (due to an external in-plane field or exchange interaction); the other is a purely orbital mechanism~\cite{Banerjee2023phase,Davydova2022} based on the finite Cooper pair momentum induced in the superconducting leads by the flux associated to the in-plane field~\cite{Banerjee2023phase}. This flux is finite if the parent superconducting film and the 2DEG are spatially separated.

The main difference between the two pictures is the expected dependence on $V_g$. Such dependence naturally emerges since $V_g$ affects the band alignment and thus the Rashba coefficient $\alpha_\mathrm{R}$. Both $\alpha_\mathrm{R}$ and the electron density $n$ critically affect $\varphi_0$, which, in turn, determines $\eta$ in multichannel systems. In contrast, the orbital mechanism~\cite{Banerjee2023phase} hardly depends on the gate voltage~\footnote{Max Geier, Karsten Flensberg, private communication.}. As discussed in the Supplementary Information, the gate voltage also affects the magnetochiral anisotropy for the inductance~\cite{Baumgartner2022}, an effect that is strictly related to the supercurrent rectification~\cite{Costa2023sign}.
The observed strong gate dependence of $\varphi_0$ and $\eta$ indicates
that the Rashba-based mechanism must certainly play an important role in the SDE. On the other hand, the orbital mechanism cannot be ruled out by our observations: it could still coexist with the spin-orbit-based mechanism.

Finally, we would like to stress that, even though we make use of a SQUID to link $\eta$ to $\varphi_0$, our point does not concern the (trivial and long known~\cite{barone}) SDE of the asymmetric SQUID as a whole. Our focus is exclusively on the intrinsic SDE in a single, homogeneous junction (JJ2).

In conclusion, we have shown the coexistence of anomalous Josephson effect and supercurrent rectification  by measuring both effects on the same Josephson junction embedded in a SQUID. The observed gate voltage and temperature dependence are compatible with a spin-orbit based picture where supercurrent rectification arises in multichannel junctions with anomalous shift $\varphi_0$ \textit{and} skewed current-phase relation. 

Josephson diodes based on $\varphi_0$-junctions are important for both fundamental research and applications. They are novel and powerful probes of symmetry breaking in 2D superconductors~\cite{Lin2022,scammell2022theory,Volkov2023} and possible probes of topological phase transitions~\cite{Legg2023}. A recent proposal~\cite{Virtanen2023}  suggested that the anomalous Josephson effect might be used in multiterminal junctions to obtain compact nonreciprocal devices as, e.g., circulators for rf-applications~\cite{Leroux2022}.

\section{Methods}

\subsection{Experimental methods}
The heterostructure is grown by molecular beam epitaxy. The full layer sequence is reported in the Supplementary Information. The most relevant layers are the topmost ones, namely, the nominally 5~nm-thick Al film at the sample surface, a 10~nm thick In$_{0.75}$Ga$_{0.25}$As layer acting as a barrier, a 7~nm InAs layer hosting the 2DEG, followed by another  In$_{0.75}$Ga$_{0.25}$As barrier of thickness 4~nm. Structures are defined by electron beam lithography. The selective etching of Al is performed by wet chemical etching using Transene D. Deep etching processes (where the 2DEG is removed altogether) is performed using a phosphoric acid-based solution.

Transport measurements are performed in a 4-point configuration using standard lock-in techniques. To determine the SQUID critical current  $I_c^+$ and $I_c^-$ as defined in the text we take $dV/dI = 6$~$\Omega$ as a threshold.

To determine the correct offset for $B_z$ we look at the symmetry of the plot of $R=dV/dI$ versus $I$ and $B_z$, see e.g., Fig.~\ref{fig:Fig1SR}\textbf{c}. Since $R(I,B_z)\approx R(-I,-B_z)$, the center of inversion symmetry of the plot allows us to determine the applied out-of-plane field which corresponds to an effective $B_z=0$. 

\subsection{Theoretical methods}

Our theoretical model, initially developed in~Refs.~\cite{Costa2023sign,costa2023microscopic}, describes the experimentally relevant system in terms of a short S--N--S Josephson junction that couples two semi-infinite $ s $-wave superconducting~(S) regions with inherently strong Rashba SOI through a thin delta-like normal-conducting~(N) link. 
Nontrivial solutions of the 2D Bogoliubov--de Gennes equation~\cite{DeGennes1989}  
    \begin{equation}
        \left[ \begin{matrix} \hat{\mathcal{H}} & \hat{\Delta}(x) \\ \hat{\Delta}^\dagger(x) & -\hat{\sigma}_y ( \hat{\mathcal{H}} )^* \hat{\sigma}_y \end{matrix}  \right] \Psi(x,y) = E \Psi(x,y) ,
    \end{equation}
with the single-electron Hamiltonian  
    \begin{multline}
        \hat{\mathcal{H}} = \left[ -\frac{\hbar^2}{2m} \left( \frac{\partial^2}{\partial x^2} + \frac{\partial^2}{\partial y^2} \right) - \mu \right] \hat{\sigma}_0 \\ 
        + \alpha_\mathrm{R} \left( k_y \hat{\sigma}_x - k_x \hat{\sigma}_y \right) \\ 
        + \left( V_0 \hat{\sigma}_0 + V_\mathrm{Z} \hat{\sigma}_y \right) d \delta(x) ,
    \end{multline}
determine the energies~$ E $ and wave functions~$ \Psi(x,y) $ of the Andreev bound states~\cite{Andreev1966,*Andreev1966alt}, which are at the heart of the coherent Cooper-pair supercurrent transport along the $ \hat{x} $-direction; 
$ \hat{\Delta}(x) $ corresponds to the $ s $-wave superconducting pairing potential that we approximate by 
    \begin{equation}
        \hat{\Delta}(x) = \Delta^*(T) \left[ \Theta(-x) + \mathrm{e}^{\mathrm{i} \varphi} \Theta(x) \right] , 
    \end{equation}
where $ \Delta^*(T) = \Delta^*(0) \tanh ( 1.74 \sqrt{T_\mathrm{c} / T - 1} ) $ is the temperature-dependent proximity-induced superconducting gap~[from the experimental data, the induced gap at zero temperature was estimated as~$ \Delta^*(0) \approx 130 \, \upmu \mathrm{eV} $ and the critical temperature as~$ T_\mathrm{c} \approx 2 \, \mathrm{K} $] and $ \varphi $ indicates the phase difference between the two superconducting regions. 
The Rashba SOI that is present throughout the whole system is parameterized by~$ \alpha_\mathrm{R} $, $ V_0 $ and $ V_\mathrm{Z} $ represent the scalar~(spin-independent) and Zeeman~(spin-dependent) potentials inside the delta-like N link of thickness~$ d $---the magnetic field causing the Zeeman splitting is thereby aligned perpendicular to the current direction~(i.e., along~$ \hat{y} $)---, $ \mu $ is the Fermi energy, $ m $ the (effective) quasiparticle mass, and $ \hat{\sigma}_0 $ and $ \hat{\sigma}_i $ refer to the $ 2 \times 2 $~identity and $ i $th Pauli spin matrix, respectively.

After determining the Andreev-state energies~$ E(\varphi) $ as a function of the superconducting phase difference~$ \varphi $, we apply the quantum-mechanical current operator to the corresponding bound-state wave functions inside the N~link to compute in the first step the Josephson CPRs~$ I(\varphi) $ and obtain in the second step the direction-dependent critical currents necessary to quantify the SDE. 
In the simultaneous presence of SOI and Zeeman interaction, the bound-state energies depend on~$ \varphi $ through~$ E(\varphi) = \Delta^* (T) f(\varphi) $, where the generic function~$ f(\varphi) $ is no longer antisymmetric with respect to~$ \varphi $, i.e., $ f(-\varphi) \neq f(\varphi) $, reflecting the broken space-inversion and time-reversal symmetries, and the therefrom resulting nontrivial $ \varphi_0 $-phase shifts.  
Note that the only impact of temperature on~$ E(\varphi) $ is an effective rescaling~(i.e., suppression with increasing temperature) of the superconducting-gap amplitude~$ \Delta^*(T) $ according to~$ \Delta^*(T) = \Delta^*(0) \tanh ( 1.74 \sqrt{T_\mathrm{c} / T - 1} ) $, whereas the qualitative shape of~$ E(\varphi) $ is not altered by temperature. 
The total Josephson current is then given by 
\begin{equation}
    I(\varphi) = \sum_E I \big( E(\varphi) ; T=0 \big) \tanh \left( \frac{E(\varphi)}{2 k_\mathrm{B} T} \right) ,
    \label{eq:tanh}
\end{equation}
where the sum ensures to account for the current contributions of all bound states~(i.e., from all transverse channels of the junction) and $ k_\mathrm{B} $ indicates the Boltzmann constant. 
The current at zero temperature can, in the simplest case, be extracted from the thermodynamic relation~\cite{Kulik1969,*Kulik1969a} 
\begin{equation}
    I \big( E(\varphi) ; T=0 \big) = -\frac{e}{\hbar} \frac{\partial E(\varphi)}{\partial \varphi} 
    \label{Eq_Current_Final} 
\end{equation}
with the positive elementary charge~$ e $. 
The major temperature effect on the Josephson current originates therefore from the suppression of the higher-harmonic contributions in~Eq.~\eqref{Eq_Current_Final} due to the tanh term in Eq.~\eqref{eq:tanh}.

The strengths of the Rashba SOI, the scalar~(barrier), and the Zeeman potentials are measured by the dimensionless parameters $ \lambda_\mathrm{SOI} = m \alpha_\mathrm{R} / (\hbar^2 k_\mathrm{F}) $, $ Z = 2m V_0 d / (\hbar^2 k_\mathrm{F}) $, and $ \lambda_\mathrm{Z} = 2m V_\mathrm{Z} d / (\hbar^2 k_\mathrm{F}) $, respectively, where $ k_\mathrm{F} = \sqrt{2m \mu} / \hbar $ refers to the Fermi wave vector~(for the experimental parameters, $ k_\mathrm{F} \approx 3 \times 10^8 \, \mathrm{m}^{-1} $). 
In agreement with our earlier studies~\cite{baumgartner2020,Costa2023sign}, we assume~$ Z=0.5 $---mimicking an average junction transparency of $ \overline{\tau} = 1/[1+(Z/2)^2] \approx 0.94 $---and $ \lambda_\mathrm{SOI} = 0.661 $---corresponding to Rashba SOI~$ \alpha_\mathrm{R} \approx 15 \, \mathrm{meV} \, \mathrm{nm} $. 
For a typical $ g $-factor of $ |g^*| \approx 12 $, the Zeeman parameter $ \lambda_\mathrm{Z} = -0.40 $ used for our theoretical calculations in the main text corresponds to the magnetic field~$ B_y \approx -100 \, \mathrm{mT} $.

\begin{acknowledgments}
We thank Max Geier, Abhishek Banerjee, and Karsten Flensberg for fruitful discussions.
    Work at Regensburg University was funded by the Deutsche Forschungsgemeinschaft (DFG, German Research Foundation) through Project-ID 314695032–--SFB 1277~(Subprojects B05, B07, and B08)---and Project-ID 454646522---Research grant ``Spin and magnetic properties of superconducting tunnel junctions''~(A.C. and J.F.). 
    D.K.~acknowledges partial support from the project 
IM-2021-26 (SUPERSPIN) funded by the Slovak Academy of Sciences via the programme IMPULZ 2021.
\end{acknowledgments}
\vspace{2mm}

\section{Author contributions}
S.R., T.A., and J.B fabricated the devices and performed initial transport characterization of the hybrid superconductor/semiconductor wafer. S.R. and T.A. performed the measurements with the SQUID device. S.R., C.S. and N.P. conceived the experiment. 
A.C., D.K., and J.F. formulated the theoretical model. A.C.~performed the numerical simulations of the temperature dependence of the SDE and $\varphi_0$. N.P., S.R., C.S., A.C., and D.K. wrote the manuscript. 
T.L., S.G., and G.C.G.~designed the heterostructure and conducted MBE growth.
S.R., T.A.~and N.P.~analyzed the data.
M.J.M.~supervised research activities at Purdue.

\section{Competing interests}		
The authors declare no competing interests.


\section{Code availability}
The computer codes that support the theoretical results, the plots within this paper and other findings of this study are available from the corresponding author upon reasonable request.


%

\clearpage
\newpage
\beginsupplement

\onecolumngrid
\begin{center}
\textbf{\Large Supplementary Information} 

\vspace{0.5cm}

\noindent \textbf{\large Link between supercurrent diode and anomalous Josephson effect revealed by gate-controlled interferometry}
\end{center}

\section{Wafer growth and sample fabrication}
The hybrid superconductor-semiconductor layer stack is shown in Fig.~\ref{fig:Fig1_Supp_SR}\textbf{a}. The active layer consists of a bottom barrier of 4~nm In$_{0.75}$Ga$_{0.25}$As, a 7~nm InAs quantum well, and a top barrier of 10~nm In$_{0.75}$Ga$_{0.25}$As. Growth of the semiconductor stack is followed by in-situ deposition of 5~nm aluminum \cite{ZhangPRM2023}.

The mesa of the SQUID device is wet etched using a solution of H$_3$PO$_4$:C$_6$H$_8$O$_7$:H$_2$O$_2$:H$_2$O (1.2:14:2:88) to a depth of 250~nm. The Josephson junctions are defined using 
standard electron beam lithography. Prior to etching of the junctions, we expose the developed structures to oxygen plasma for 5~s in order to remove residues of PMMA resist. Selective etching of the aluminum in the junctions is then performed with aluminum etchant type D from Transene at 50~$^\circ$C for 3~s. Both junctions are covered with 60~nm aluminum oxide grown by atomic layer deposition at 80~$^\circ$C. Before the first pulse (trimethylaluminum), the sample is dried under nitrogen flow for 2~hours. Finally, a Ti/Au (5~nm / 100~nm) top-gate is deposited onto JJ2 by electron beam evaporation.

\section{Basic characterisation of the quantum well and the superconductor}
\begin{figure*}[tb]
\centering
\includegraphics[width=\textwidth]{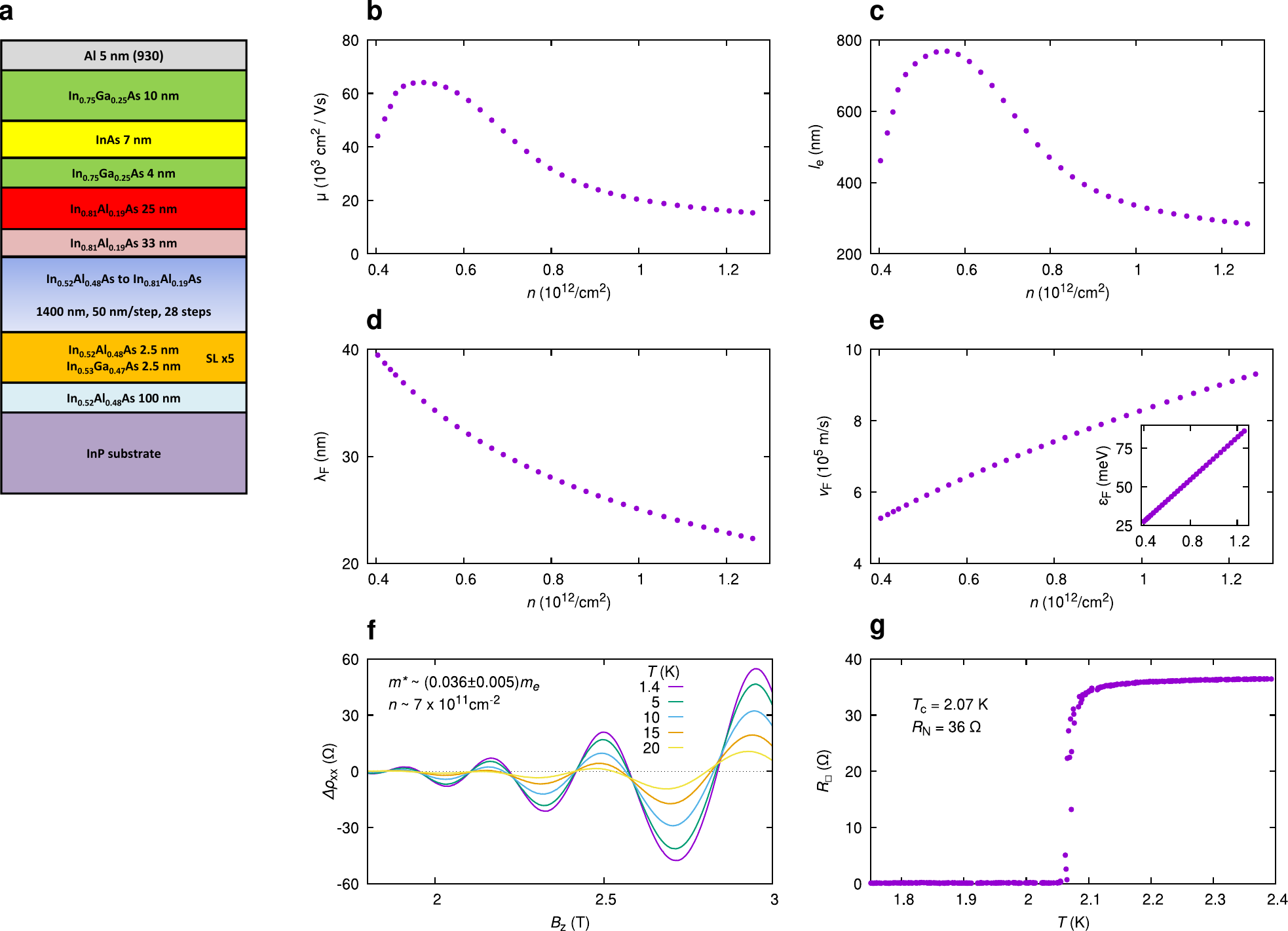}
\caption{
\textbf{a}, Layer stack of the used wafer.
\textbf{b-e}, Transport properties of the quantum well measured with a top gated Hall bar at temperature $T = 1.5$~K. The inset of panel \textbf{e} shows the Fermi level $\varepsilon_F$ versus $n$. The range in gate voltage corresponding to the $n$-range displayed in panels \textbf{b-e} is [-3.5~V,-0.2~V].
\textbf{f}, Amplitude of Shubnikov-de Haas oscillations for temperatures between $1.4$~K and $20$~K.
\textbf{g} $R(T)$ Characteristic of a Hall bar device with the pristine aluminum film.
}

\label{fig:Fig1_Supp_SR}
\end{figure*}

Basic transport properties of the semiconductor quantum well are performed with a standard Hall bar device. The aluminum film is removed from the Hall bar and a top-gate is fabricated using the methods described above.
In Figures~\ref{fig:Fig1_Supp_SR}\textbf{b-e} we show Hall mobility, mean free path, Fermi wavelength, and Fermi velocity as a function of carrier density at $T = 1.5$~K. The peak mobility $\mu >60000$~cm$^2/$Vs is obtained at the density $n \sim 5 \times 10^{11}$~cm$^{-2}$.

The electron effective mass is determined from the temperature dependence of Shubnikov-de Haas (SdH) oscillations, following \cite{Yuan2020}. Figure~\ref{fig:Fig1_Supp_SR}\textbf{f} shows SdH oscillations in the resistance with a second-order polynomial background removed in order to obtain the amplitude of each oscillation as a function of temperature. The carrier density obtained from the oscillations is $\sim 7 \times 10^{11}$~cm$^{-2}$. For each maximum and minimum of the oscillations we obtain an effective mass by fitting the temperature dependence of the amplitude with the formula
\begin{equation}
    A_\text{SdH}(T) \sim \frac{2\pi^2 k_B T / \hbar \omega_c}{\sinh(2\pi^2 k_B T / \hbar \omega_c)}
\end{equation}
The obtained mass $m^*/m_e = 0.036 \pm 0.005$ is slightly lower compared to the value $m^*/m_e = 0.04$ found in \cite{Yuan2020}.

Figures~\ref{fig:Fig1_Supp_SR}\textbf{g} shows the resistance of the heterostructure, including the aluminum film, as a function of temperature, measured in a Hall bar geometry.
Due to the low thickness of the film, the critical temperature is enhanced to $2.07$~K with a normal state sheet resistance of 36~$\Omega$.

\section{Measurement techniques and data evaluation}
Measurements are performed in a dilution cryostat with a base temperature of $40$~mK.
Differential resistance $dV/dI(I_{DC})$ is measured in a standard 4-point setup using a current excitation of $I_\text{ac} = 10$~nA(rms) at frequency $f = 2.333$~kHz. The voltage is measured using a differential bipolar junction preamplifier (SR552) and a digital lock-in amplifier (SR830). All measurement lines are filtered at the head of the cryostat using LC filters (Tusonix 4201-053). Below the mixing chamber we use highly resistive CuNi/CuNi coaxial lines (GVLZ185) with a large attenuation above $100$~MHz. The fields $B_z$ and $B_y$ are applied using a 2-axis vector magnet, and the rotation of the in-plane field is performed by mounting the sample on an Attocube piezo rotator (ANRv220).
A resolution below $50$~nT in $B_z$ is realized by exchanging the magnet power supply with a high precision current source (Keysight B2961B).
Before starting a measurement at a new configuration of in-plane field and/or rotation angle we heat our sample above the critical temperature $T_c \sim 2K$ and perform a field cooling in zero out-of-plane field.
The gate voltage $V_g$ is applied using a source meter (Keysight B2901B). For gate voltages in the range $V_g = -2.5$~V$\ldots 1$~V the leakage current is below the resolution of the instrument ($|I_\text{leak}| < 1$~pA). For the definition of the SQUID critical current we use a threshold resistivity of $dV/dI = 6$~$\Omega$. Using a different definition for the threshold has no visible effect on the results. 

For the determination of $\varphi_0$ and $\eta$ the SQUID critical current $I_c(B_z)$ is measured for the sequence of gate voltages $V_g = -2.5$~V, $-1.5$~V, ..., $0.5$~V, $1$~V.
The accuracy is improved by repeatedly cycling the gate voltage through this sequence (40 cycles for most measurements). After obtaining $\varphi_0$ and $\eta$ we average over all cycles. The error bars in Fig.~2\textbf{e} for the values of $\eta$ are found as the standard error of the mean.

\section{Modeling of inductive screening effects in the asymmetric SQUID}
\begin{figure*}[tb]
\centering
\includegraphics[width=\textwidth]{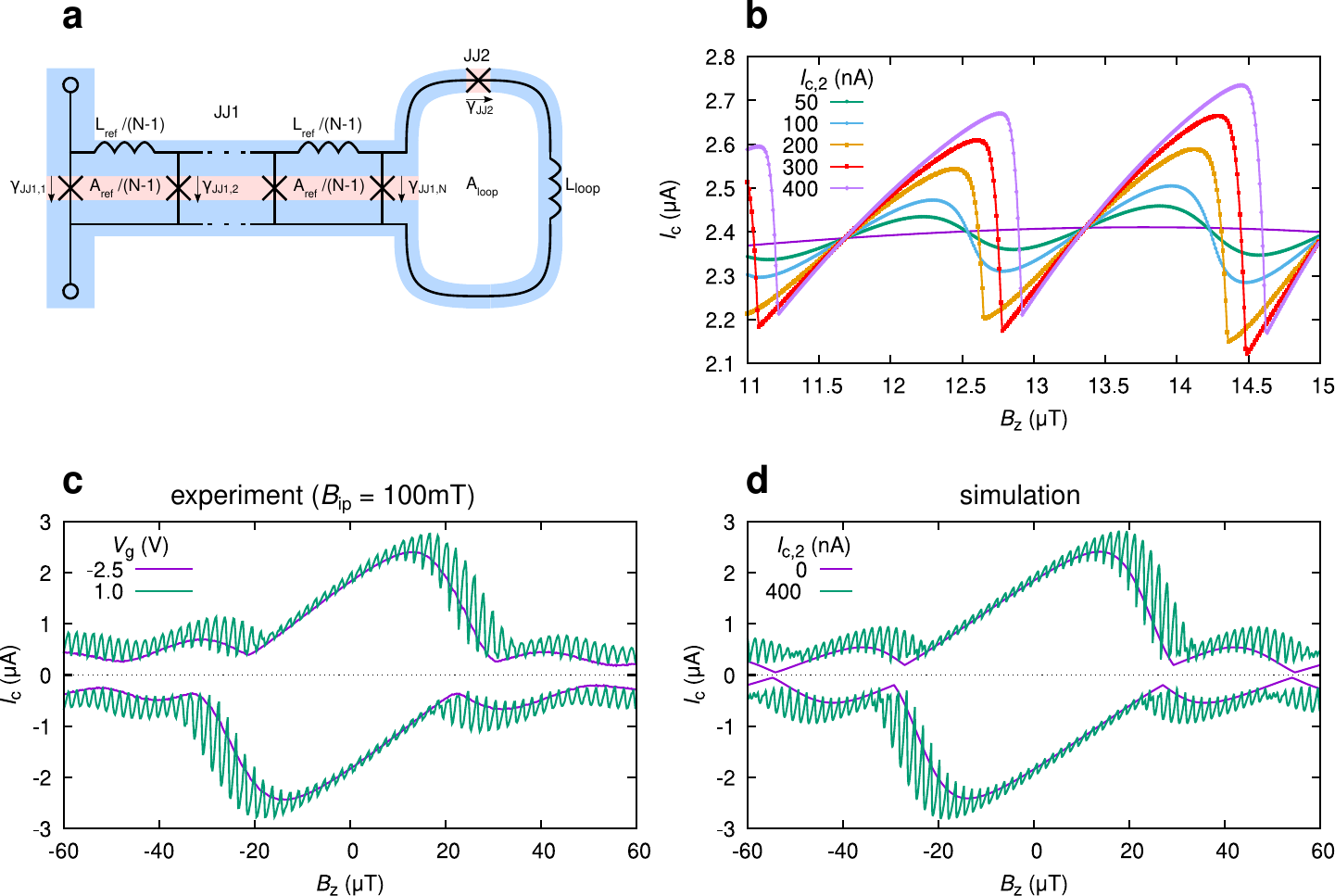}
\caption{\textbf{a}, Minimal circuit model of the SQUID device. \textbf{b,d}, Simulated critical currents of the SQUID with parameters as described in the text. \textbf{c}, experimental values of positive and negative critical currents  for $B_\text{ip} = 100$~mT and $\theta = -90$~$^\circ$.}
\label{fig:Fig2_Supp_SR}
\end{figure*}

The obtained critical currents of the SQUID device are strongly affected by the large kinetic inductance of the aluminum electrodes. 
A minimal circuit model of our device, which includes the inductance of the electrodes of JJ1 and the loop, is show in Figure~\ref{fig:Fig2_Supp_SR}\textbf{a}. The reference junction JJ1 is modeled as a network of $N$ JJs in parallel, each with critical current $I_{c,1}/N$. Setting $N=20$ is sufficient to model the central lobe of the Fraunhofer interference pattern of JJ1. The effective area of JJ1 is $A_\text{ref}$, while the area of the large loop is $A_\text{loop}$. 
The positive and negative critical currents for a given out-of-plane field $B_z$ can be calculated as a function of the phase difference $\gamma_\text{JJ2}$ over JJ2. Given $\gamma_\text{JJ2}$ the phase difference $\gamma_\text{JJ1,N}$ over the rightmost junction in JJ1 is
\begin{equation}
    \gamma_{\text{JJ1},N} = \gamma_\text{JJ2} - 2\pi (B_z A_\text{loop} - L_\text{loop} I_{c,2}\sin(\gamma_\text{JJ2}))/ \Phi_0
\end{equation}
The phase differences $\gamma_{\text{JJ1,}i}$ are found by iteration:
\begin{equation}
    \gamma_{\text{JJ1,}i} = \gamma_{\text{JJ1},i} - 2\pi (B_z A_\text{ref}/(N-1) - L_\text{ref}/(N-1)I_{i+1})/\Phi_0
\end{equation}
where $I_i = I_{i+1} + I_{c,1}/N\sin(\gamma_{\text{JJ1,}i+1})$ is the current flowing through the inductor between junctions $i$ and $i+1$. 
The positive and negative critical currents are then found as the extremal values of $I_0(\gamma_\text{JJ2})$.
In Figures\ref{fig:Fig2_Supp_SR}\textbf{b-d} we model the SQUID critical current obtained for a in-plane-field $|B_y| = 100$~mT. A good match to the data can be found with the following parameters: $A_\text{loop} = 1150$~(\textmu m)$^2$, $I_{c,1} = 2.48$~$\mu$A, $A_\text{ref} = 70$~(\textmu m)$^2$, $L_\text{ref} = 0.8$~nH, and $L_\text{loop} = 1.5$~nH. From the value of $L_\text{loop}$ we determine the sheet inductance of the aluminum film $L_\square = L_\text{loop} / N_{\square,\text{loop}} = 1.5$~nH$/46 \sim 30$~pH.
The kinetic inductance of the loop largely exceeds the geometric inductance which we estimate as $L_\text{loop,geo} \lesssim 100$~pH.

The amplitude of the SQUID oscillations at $V_g = 1$~V, $By=-100$~mT and $T=40$~mK is matched by setting $I_{c,2} = 400$~nA.
Figure~\ref{fig:Fig2_Supp_SR}\textbf{b} shows the SQUID critical current for different values of $I_{c,2}$ near the maximum of the SQUID critical current. Importantly, the crossing point with the reference line ($I_{c,2} = 0$) is the same for all values of $I_{c,2}$, showing that screening has no effect on our evaluation of $\Delta \varphi_0$.

Figure~\ref{fig:Fig2_Supp_SR}\textbf{c} shows the results of SQUID oscillation measurements performed for $V_g=1.0$~V (green) and $V_g=-2.5$~V (purple), at $B_y=-100$~mT and $T=40$~mK. A comparison with the results of our calculations in Fig.~\ref{fig:Fig2_Supp_SR}\textbf{d} makes it evident that our simple model described above correctly captures the behavior of the SQUID. We want to stress that our asymmetric SQUID displays a trivial supercurrent diode behavior (in Fig.~\ref{fig:Fig2_Supp_SR}\textbf{c,d} in general $I_c^+(B_z) \neq I_c^-(B_z)$). This is a long known feature of asymmetric SQUIDs and it is not the focus of our work. Instead, our focus is on the physics of a single, homogeneous Josephson junction (JJ2).

From the analysis of the inductive screening effects discussed above, we can conclude that:
\begin{itemize}
    \item Owing to screening effects, the SQUID oscillations do not reproduce the CPR of JJ2. However, for positive bias, one can still deduce $I_{c,2}^+$ from the maximum of the oscillations, since these two quantities are proportional, see Fig.~\ref{fig:Fig2_Supp_SR}\textbf{d}. Instead, \textbf{for positive bias, one cannot deduce $I_{c,2}^-$ from the minima of the SQUID oscillations}.
    The correct way to deduce $I_{c,2}^-$ is to look at oscillations for negative current bias. $I_{c,2}^-$ is proportional to the \textbf{minima} of the SQUID oscillations for \textbf{negative bias}, see Fig.~2\textbf{c} of the main text.
    \item As it is clearly visible in Figures~\ref{fig:Fig2_Supp_SR}\textbf{c,d}, the amplitudes of the fast SQUID oscillations are not constant. This is caused by a coupling of the loop current $I_\text{JJ2}$ to the flux of JJ1. The flux in JJ1 is changed by $\Phi \sim L_\text{ref}I_\text{JJ2}$, modulating the critical current of JJ1. For the measurement of critical currents, demonstrated in Figure~2 of the main text, we find that the critical currents for different gate voltages are only proportional to $I_{c,2}(V_g)$, with the proportionality constant depending on the chosen SQUID oscillation (i.e., on the exact $B_z$ value). As positive and negative critical currents need to be measured at opposite current bias (see previous points), the slightly different scale factors result in offsets for the extracted diode efficiency $\eta = 2(I_c^+ - |I_c^-|) / (I_c^+ + |I_c^-|)$. We note that the gate slope $d \eta / d V_g$ is nevertheless a well defined quantity and not affected by the screening effects.
    \item As a consequence of the $B_z$-dependence of the amplitude of the SQUID oscillations, the correct way to compare an oscillation with positive bias with one with negative bias is to consider oscillations with opposite $B_z$, namely oscillations which are inversion-symmetric around the origin of the $I_c(B_z)$ graph, see arrows in Fig.~2\textbf{a} of the main text. Indeed, by the global time-reversal symmetry $I_c^+(B_z)=-I_c^-(-B_z)$ in absence of in-plane field. Taking precisely opposite oscillations, as described in the previous point, should make screening effect irrelevant, so that, in principle, the observation of finite $\eta$ for finite $B_{\text{ip}}$ should be considered as a real effect and not an artifact of screening. However, the combination of $\varphi_0$-shift and the strong dependence of the SQUID oscillations on $B_z$ makes it possible to measure (following the aforementioned procedure) a finite $\eta$ even in absence of a genuine SDE in JJ2. In fact, owing to the finite $\varphi_0$, opposite SQUID oscillations do not occur anymore at exactly opposite $B_z$, and the strong $B_z$ dependence of the SQUID oscillations can render the opposite oscillation amplitudes not exactly equal even in absence of intrinsic SDE.  The effect is very small, still we maintained a conservative approach and decided to look at $\partial \eta /\partial V_g$ instead of $\eta$ itself. Also, we can exclude that the gate dependence of $\eta$ is a simple artifact of that of $\varphi_0$ (due to the mechanism just described) for the following reasons: (i) from our calculations the measured $\eta$ is too large to be entirely generated by the mechanism just described ($\varphi_0$ plus $B_z$-dependent oscillations). (ii) The gate dependence of $\varphi_0$ and $\eta$ are different: the former is nonlinear, the latter is linear (compare Fig.~1\textbf{g} and Fig.~2\textbf{e} of the main text). (iii) The temperature dependence of $\Delta \varphi_0$ and that of $\partial \eta /\partial V_g$ is dramatically different, see Fig.~3 of the main text.
    \item From Fig.~\ref{fig:Fig2_Supp_SR}\textbf{b} we can conclude that inductive screening  does not affect the value of $B_z$ at which $I_2=0$ (the curves all cross the baseline at the same point). Clearly, for positive bias, it is important to only consider the crossing with the baseline \textit{with positive slope}, which is the significant one. This explains our operational definition of $B_0$, which is important to determine $\varphi_0$. 
    \item The inductive screening effects depend on the kinetic inductance of the Al film and on the magnitude of the supercurrent flowing the in the SQUID arms. By increasing $B_{\text{ip}}$ in the range under study, the decrease of the critical current $I_c$ is more important than the increase of the kinetic inductance of the arms.  Therefore, the larger $B_{\text{ip}}$, the smaller the impact of screening. Since $\varphi_0$ measurements are more sensitive to inductive screening, we were able to extract $\Delta \varphi_0$ only for $B_{\text{ip}} \geq 100$~mT.  
\end{itemize}

\section{Gate dependence of the magnetochiral anisotropy for the inductance}

In this section we comment about the gate dependence of the magnetochiral anisotropy (MCA) of the inductance, which we measured in our previous work, Ref.~\cite{Baumgartner2022}. In that work, we established an analogy between the known MCA for the resistance (e.g. in the fluctuation regime of a noncentrosymmetric superconductor) and MCA for the inductance. This latter was the new quantity introduced in Ref.~\cite{Baumgartner2022}, where we showed that the Josephson inductance can be written as 
\begin{equation}
    L=L_0[1+\gamma_L\hat{e}_z\cdot(\vec{B}\times \vec{I})],
    \label{eq:MCAind}
\end{equation}
where $\gamma_L \equiv 2 L^{\prime}_0/(L_0B_{ip})$, with $L_0^{\prime} \equiv \frac{dL}{dI}|_{I=0}$. 
Equation~\ref{eq:MCAind} mirrors the Rikken formula~\cite{RikkenPRL2001}
\begin{equation}
    R=R_0[1+\gamma\hat{e}_z\cdot(\vec{B}\times \vec{I})],
    \label{eq:MCAres}
\end{equation}
which holds for noncentrosymmetric conductors with nonreciprocal resistance, for which $\gamma$ is defined as $ 2 R^{\prime}_0/(R_0B_{ip})$, with $R_0^{\prime} \equiv \frac{dR}{dI}|_{I=0}$. 

It must be kept in mind that both the MCA coefficient for the resistance ($\gamma$) and that for the inductance ($\gamma_L$) are dimensional quantities which scale with the inverse of the sample width in quasi 2D conductors. In Ref.~\cite{Baumgartner2022}, we have plotted in Fig.~2d the gate dependence $\gamma_L$. At first glance, it might appear that no dependence is observed (the curves in Fig.~2d of Ref.~\cite{Baumgartner2022} fall on top of each other). However, it must be kept in mind that $\gamma_L$ trivially depends on the number of channels $N$ (or, equivalently,  on the width). Since $I\propto N$ and  $L_0 \propto N^{-1}$, then $\gamma_L$ must scale as $1/N$. Therefore, in Ref.~\cite{Baumgartner2022} the fact that $\gamma_L$ appears constant is due to the fact that the expected increase of the specific magnetochiral character (considered ``per channel'') is compensated by the decrease due to the trivial increase of $N$ (decrease of $N^{-1}$).

\begin{figure*}[tb]
\centering
\includegraphics[width=\textwidth]{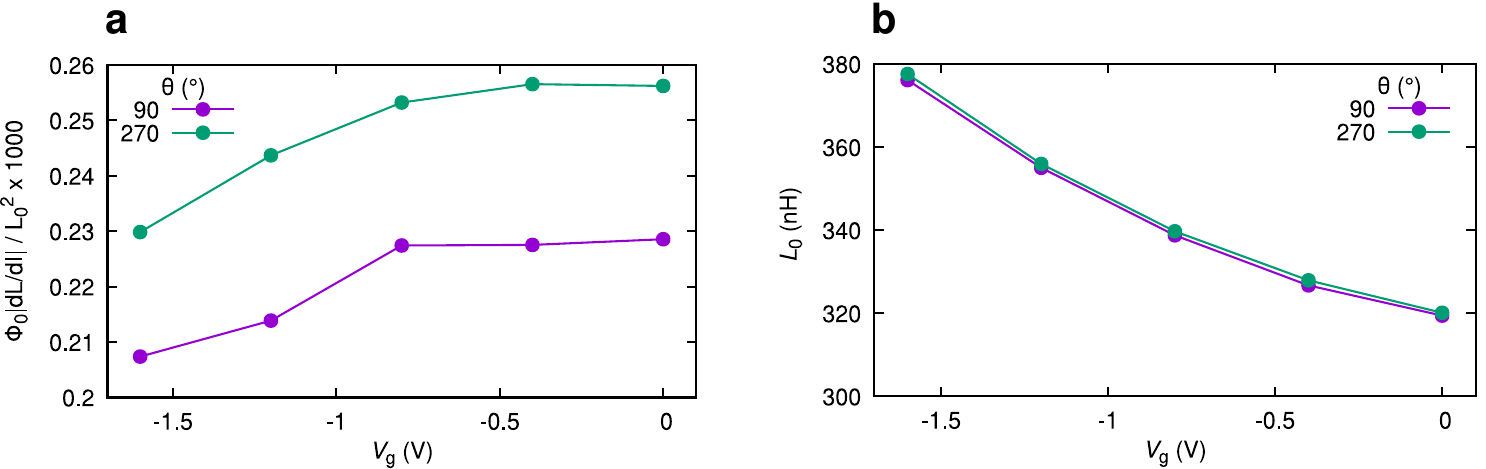}
\caption{Magnetochiral anisotropy of the Josephson inductance for 1D array at $B_\text{ip} = 100$~mT.}
\label{fig:Fig3_Supp_SR}
\end{figure*}

The best way to represent the gate-induced enhancement of the MCA for the inductance (which mirrors the increase in $|\varphi_0|$ and $\eta$ with $V_g$ demonstrated in the present article) is to represent the MCA with a quantity which does not depend on $N$. To this end, we plot in Fig.~\ref{fig:Fig3_Supp_SR}\textbf{a} the $V_g$ dependence of the quantity $\Phi_0 |L_0^{\prime}|/L_0^2$ for the same data set of Fig.~2d in Ref.~\cite{Baumgartner2022}. Data refer to $B_\text{ip}=100$~mT and $\theta=90^{\circ}$ (purple) and $\theta=270^{\circ}$ (green). We clearly observe an increase in the magnetochiral character with $V_g$. The increase is relatively small due to the fact that in that sample the gate had a much weaker effect compared to the sample studied in the present article. To give an idea of this, in Fig.~\ref{fig:Fig3_Supp_SR}\textbf{b} we plot $L_0$ as a function of $V_g$, showing that sweeping $V_g$ from -1.5 to 0~V corresponds to a change in $L_0$ of less than 20\%. 

\section{Angle dependence of the anomalous phase shift}
\begin{figure*}[tb]
\centering
\includegraphics[width=\textwidth]{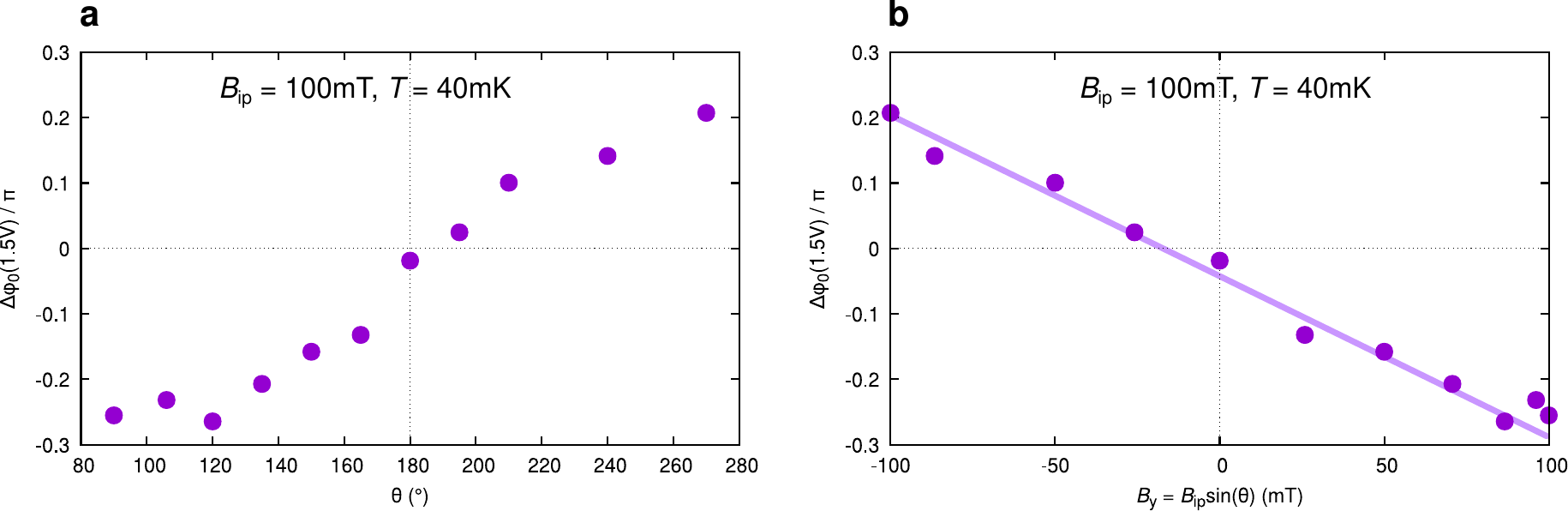}
\caption{\textbf{a}, Anomalous phase shift for $B_{ip} = 100$~mT as a function of $\theta$ as defined in the main text.
\textbf{b}, Anomalous phase shift as a function of $B_y = B_{ip}\sin(\theta)$.
}
\label{fig:Fig4_Supp_SR}
\end{figure*}

Figure~\ref{fig:Fig4_Supp_SR}\textbf{a} shows the dependence of the anomalous phase difference $\Delta\varphi_0(V_g = 1.5$~V$, B_\text{ip} = 100$~mT$)$ on the rotation angle $\theta$ defined in the main text. The observed behaviour is in good agreement with the expected relation $\varphi_0 \sim \sin(\theta)$.
In Fig. \ref{fig:Fig4_Supp_SR}\textbf{b} we plot $\Delta\varphi_0$ as a function of the $B_y$ component of the in-plane field ($B_y := B_{ip}\sin(\theta)$), showing a good agreement with the expected linear behaviour.

\section{Supplemental data on temperature dependence}
\begin{figure*}[tb]
\centering
\includegraphics[width=\textwidth]{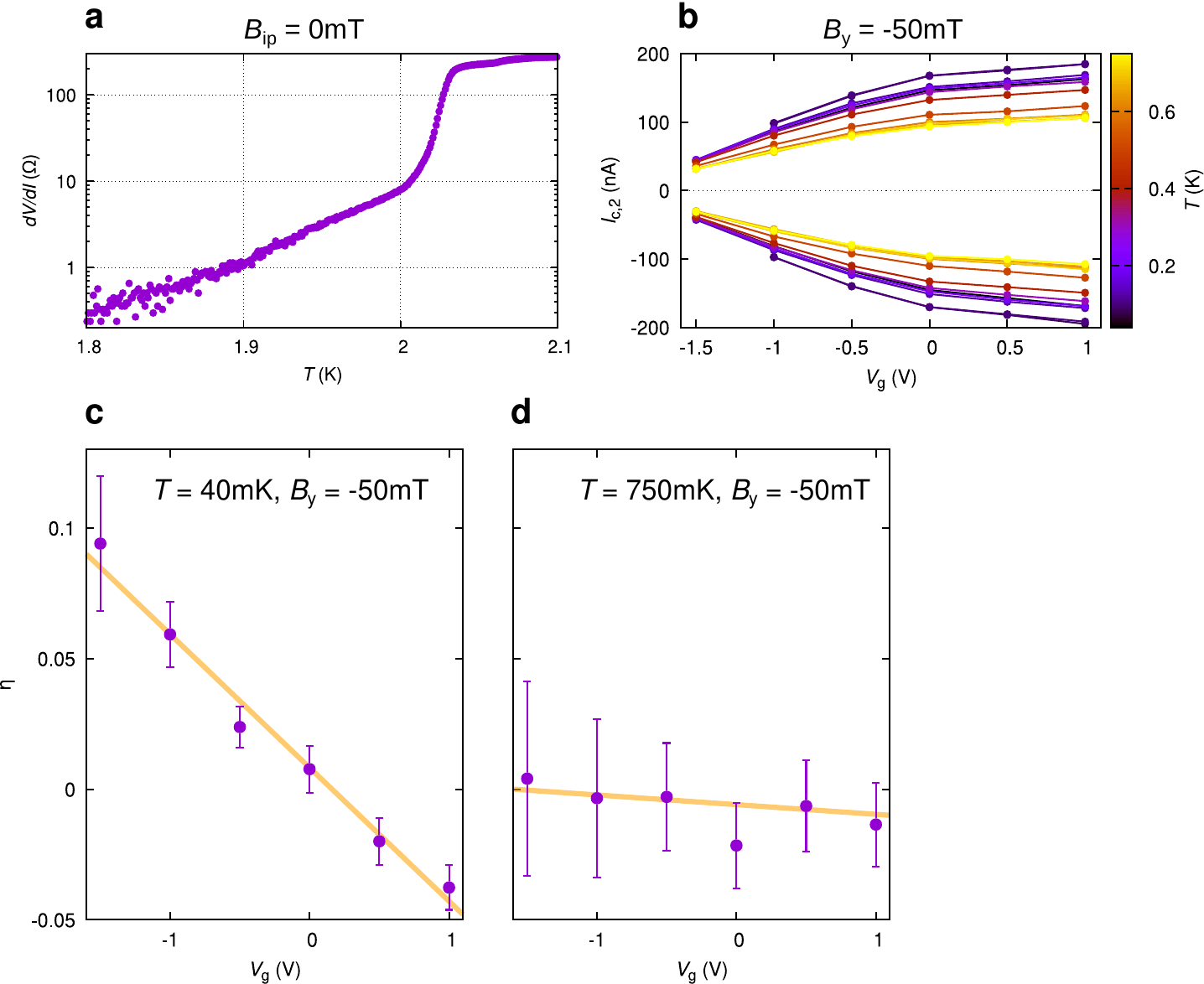}
\caption{\textbf{a}, $dV/dI(T)$ of the SQUID device at zero DC bias and zero magnetic field. \textbf{b}, Critical current $I_{c,2}^\pm(V_g)$ at $B_y = -50$~mT for different temperatures. \textbf{c} and \textbf{d}, Diode efficiency $\eta(V_g)$ for $T = 40$~mK and $T = 750$~mK.  
}
\label{fig:Fig5_Supp_SR}
\end{figure*}

Figure~\ref{fig:Fig5_Supp_SR}\textbf{a} shows the temperature dependence of the differential resistance $dV/dI$ of the SQUID for zero DC bias current and zero magnetic field. 
The positive and negative critical currents $I_{c,2}^\pm$ of JJ2 corresponding to Fig. 3\textbf{c} and \textbf{d} of the main text are shown in Fig. \ref{fig:Fig5_Supp_SR}\textbf{b}. At $T=750$~mK the critical current is reduced by $\sim 50$~\%.
Fig. \ref{fig:Fig5_Supp_SR}\textbf{c} and \textbf{d} show $\eta(V_g)$ for the temperatures $T = 40$~mK and $T = 750$~mK (corresponding to the first and last data points in Fig. 3\textbf{d} of the main text). At $T = 750$~mK we observe larger statistical errors in the determination of the critical currents, leading to the larger error bars. By increasing the temperature, the superconducting-to-normal transition in the differential resistance becomes less sharp, leading to the increased noise in the determination of critical currents, and thereby the diode efficiency.

\end{document}